\documentclass[article,10pt,a5paper,oneside,cyremdash]{ncc}
\FromMargins[h]{2.4cm}{1.6cm}{.8cm}{1.6cm}
\usepackage{cmap}
\usepackage[boldsans]{concmath}
\usepackage{amssymb,fancyvrb,cite,icomma,indentfirst}
\fvset{fontsize=\footnotesize,frame=single,framesep=.4em,
       xleftmargin=0.2em,xrightmargin=0.2em}
\usepackage[utf8]{inputenc}
\usepackage[T1,T2A]{fontenc}
\usepackage[english,russian]{babel}

\usepackage[pdftex,unicode]{hyperref}
\hypersetup{unicode=true,
    pdfauthor={P.V. Moskalev, K.V. Grebennikov, V.V. Shitov},
    pdfsubject={Interval estimation of the mass fractal dimension for isotropic sampling percolation clusters},
    pdftitle={P.V. Moskalev et al. Interval estimation of the mass fractal dimension for isotropic sampling percolation clusters},
    pdfkeywords={mathematical modeling, Monte Carlo methods, site percolation, percolation cluster, mass fractal dimension} }

\begin{document}

\begin{titlepage}

\English
\begin{flushleft}\bfseries
UDC 519.676
\end{flushleft}
\begin{flushleft}\bfseries\Large
Interval estimation of the mass fractal dimension for isotropic sampling percolation clusters
\end{flushleft}
\begin{flushleft}\bfseries
P.V. Moskalev (Voronezh, moskalefff@gmail.com), \\
K.V. Grebennikov, V.V. Shitov (Voronezh, svw@list.ru)
\end{flushleft}
\begin{flushleft}
IV International scientific conference 
<<Modern problems of applied mathematics, control theory and mathematical modeling>>, \\
Voronezh, 12\--17 September 2011
\end{flushleft}
\begin{flushleft}
Received: July 14, 2011
\end{flushleft}

\paragraph*{Abstract.} 
This report focuses on the dependencies for the center and radius of the confidence interval that arise when estimating the mass fractal dimensions of isotropic sampling clusters in the site percolation model.
\vspace{1ex}

\noindent\rule{\textwidth}{.5pt}

\Russian
\begin{flushleft}\bfseries
УДК 519.676
\end{flushleft}
\begin{flushleft}\bfseries\Large
Интервальное оценивание массовой фрактальной размерности для выборки изотропных перколяционных кластеров
\end{flushleft}
\begin{flushleft}\bfseries
П.В. Москалев (Воронеж, moskalefff@gmail.com), \\
К.В. Гребенников, В.В. Шитов (Воронеж, svw@list.ru)
\end{flushleft}
\begin{flushleft}
IV Международная научная конференция 
<<Современные проблемы прикладной математики, теории управления и математического моделирования>>, \\
Воронеж, 12\--17 сентября 2011 г.
\end{flushleft}
\begin{flushleft}
Поступил в Оргкомитет: 14 июля 2011 г.
\end{flushleft}

\paragraph*{Аннотация.}
Настоящий доклад посвящён зависимостям для центра и радиуса доверительного интервала, возникающего при оценивании массовой фрактальной размерности выборки изотропных кластеров в математической модели перколяции узлов.

\normalsize
\end{titlepage}

Одной из типичных задач, возникающих при применении теории перколяции в прикладных задачах математического моделирования, является статистическая оценка параметров перколяционного кластера, представляющего собой реализацию псевдослучайного процесса, в общем случае зависящего от типа перколяционной решётки, её размера, доли достижимых узлов и/или связей, стартового подмножества узлов и распределения взвешивающей последовательности псевдослучайных чисел. В наших недавних работах \cite{moskaleff.2011.01, moskaleff.2011.02} описана новая эффективная методика, применимая для получения репрезентативной статистической оценки массовой фрактальной размерности кластера в широком диапазоне долей достижимых узлов $p$ изотропной перколяционной решётки. Методика основана на построении линейной регрессии для выборочных векторов логарифмов не абсолютных суммарных частот $\ln n_i$, а относительных суммарных частот $\ln v_i$ узлов кластера, покрываемых квадратами текущего размера $\ln r_i$: 
$\ln v_i = d_{b1} \ln r_i + d_{b0} + e_{bi}$, где сумма квадратов отклонений $e_{bi}$ минимизируется методом наименьших квадратов.

Это вполне правомерно, так как средние значения абсолютных частот $\overline{\vphantom{i}n}_i$ узлов перколяционного кластера пропорциональны их относительным частотам $v_i$ в данной выборке, что означает равенство математических ожиданий оценок коэффициентов регрессии $d_{b1}$ и $d_{c1}$, найденных по новой и классической методикам. По новому методу процессы построения выборочных векторов $(\ln r_i, \ln v_i)$ и оценивания коэффициента регрессии $d_{b1}$ разделяются, что не только сокращает общую трудоёмкость вычислений, но и существенно расширяет диапазон докритических долей достижимых узлов решётки $p < p_c$, для которых возможна оценка массовой фрактальной размерности кластеров $d_{b1}$. Последнее обстоятельство особенно существенно, так как в докритической области значений $p$ доля реализаций, пригодная для нахождения репрезентативной оценки массовой фрактальной размерности классическим методом, быстро падает, порождая чрезмерное увеличение радиуса доверительного интервала. 

\begin{figure}[hbt]
\centering
\includegraphics[width=.47\linewidth]{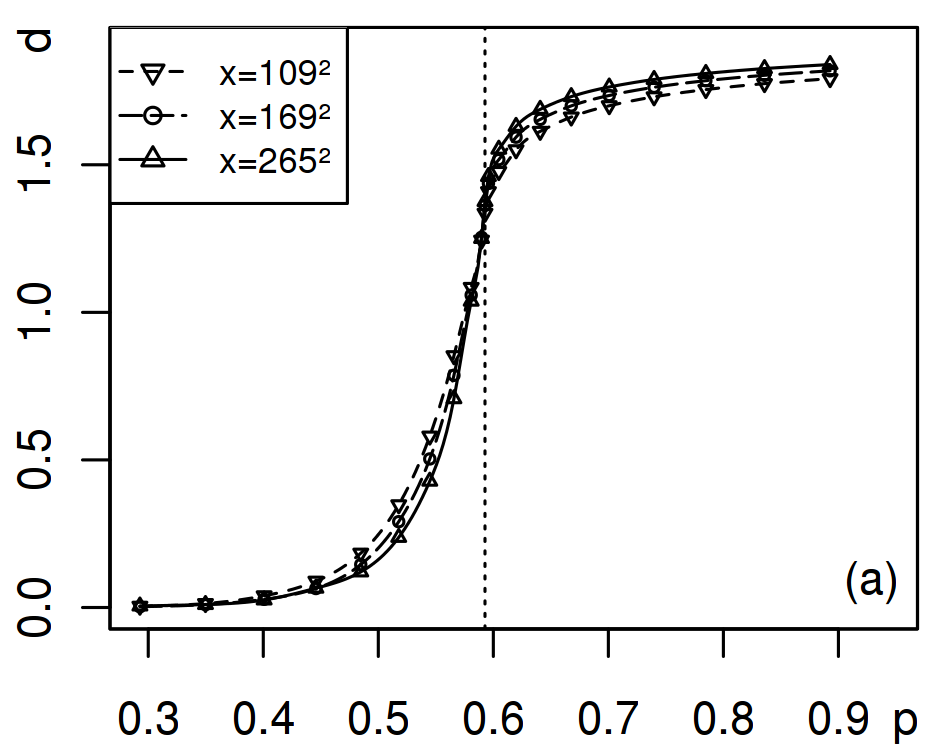}\quad
\includegraphics[width=.47\linewidth]{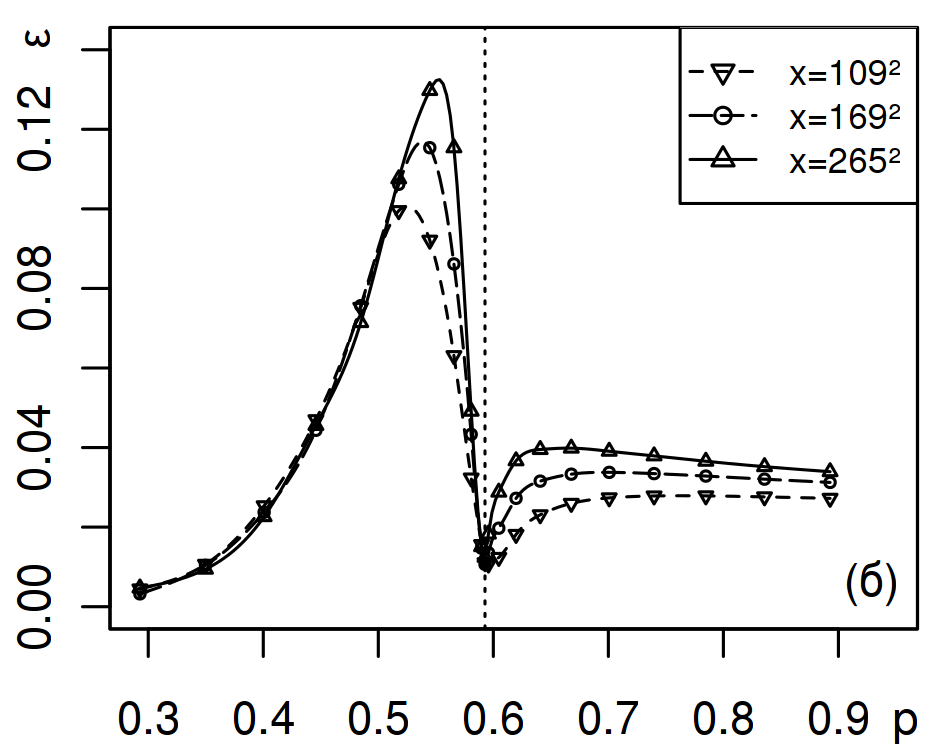}
\caption{Зависимости для центра $d$ и радиуса $\varepsilon$ доверительного интервала массовой фрактальной размерности $\mathbf{I}_{0,95}(d_{b1}) = d \pm \varepsilon$ от доли достижимых узлов $p$ при размерах квадратной решётки $x = 109^2, 169^2, 265^2$ узлов: а)~график $d(p|x)$; б)~график $\varepsilon(p|x)$}
\label{pic:pdex}
\end{figure}

На рис.\,\ref{pic:pdex} показано изменение интервальной оценки массовой фрактальной размерности $\mathbf{I}_{0,95}(d_{b1}) = d \pm \varepsilon$, от доли достижимых узлов $p$ при различных размерах квадратной решётки $x$. Каждая точка на графиках $d(p|x)$ и $\varepsilon(p|x)$ была найдена по выборочной совокупности объёмом 999 реализаций кластеров, полученных в результате статистического моделирования задачи узлов со стартовым подмножеством в центре изотропной квадратной решётки. Регрессионная модель строилась по 9 точкам, равномерно распределённым в логарифмическом масштабе $\ln r_i$. Вертикальная пунктирная линия соответствует критическому значению доли достижимых узлов $p_c = 0,592745\ldots$, называемому также порогом перколяции. С учётом того, что зависимость точечной оценки массовой фрактальной размерности от доли достижимых узлов $d(p)$ имеет вид, близкий к логистической функции, а наибольшее значения её градиента приходятся на околокритическую область $p\approx p_c$, по оси абсцисс было принято квадратичное распределение расчётных точек: $p_k = p_c \pm 0,003\,k^2$, где $k = 0, 1, \ldots, 10$.

Анализ графика $\varepsilon(p|x)$ показывает, что радиус доверительного интервала достигает глобального максимума при докритических значениях абсциссы $p_1 < p_c$, причём ордината этого максимума $\varepsilon_1$ растёт вместе с размером решётки $x$, а отклонение его абсциссы от критического значения $(p_c-p_1)$\--- падает. При дальнейшем стремлении $p$ к критическому значению функция $\varepsilon(p|x)$ достигает локального минимума, а при сверхкритических значениях абсциссы $p_2 > p_c$\--- локального максимума, ордината которого $\varepsilon_2$ также растёт вместе с размером решётки $x$, а отклонение его абсциссы от критического значения $(p_2-p_c)$\--- падает. Заметим, что при $p\to 0$ и $p\to p_c$ значимость вариаций от размера решётки $x$ как для центра $d(p|x)$, так и для радиуса доверительного интервала $\varepsilon(p|x)$ массовой фрактальной размерности $d_{b1}$ быстро падает и в пределе при $n\to\infty$ зависимость $d_{b1}(p)$ приобретает кусочно\-/постоянный характер вида: $d_{b1} = 0$ при $p < p_c$; $d_{b1} = 1,358\pm 0,011$ при $p = p_c$; $d_{b1} = 2$ при $p > p_c$. 

На первый взгляд возрастающая зависимость для максимальной ординаты радиуса доверительного интервала от размера решётки $\varepsilon_1(x)$ контринтуитивна, но на самом деле она хорошо согласуется с критическим характером перколяционного процесса \cite{sahimi.1994.01}. Причина указанного поведения заключается в том, что в процессах перколяции малое приращение доли достижимых узлов в докритической области $p < p_c$ способно вызывать существенное увеличение размера для отдельных реализаций кластеров, не оказывая влияния на остальные, что и приводит к росту выборочной дисперсии. В околокритической области, при $p\approx p_c$, доля кластеров, размер которых оказывается сопоставимым с размером системы, быстро возрастает, приводя к снижению выборочной дисперсии. При этом рост дисперсии при $p\to p_1$ оказывается зависимым от размера перколяционной решётки $x$, а её падение при $p\to p_c$\--- нет, что и объясняет наблюдаемые эффекты.

На основании вышеизложенного можно сделать вывод о том, что при статистическом оценивании массовой фрактальной размерности по выборочной совокупности перколяционных кластеров в изотропной задаче узлов на квадратной решётке по предлагаемой в работах \cite{moskaleff.2011.01, moskaleff.2011.02} методике оказывается возможным одновременное повышение точности оценки и снижение вычислительной сложности задачи за счёт сокращения размера решётки $x$ до уровня, достаточного для построения адекватной модели линейной регрессии, погрешность которой и становится в данном случае определяющей.

\section*{Список литературы}

\clearpage

\English
\section*{About the authors}

\begin{description}
\item[Moskalev Pavel Valentinovich:] Ph. D., Associate Professor, Department of Mathematics and Theoretical Mechanics, Voronezh State Agricultural University after K.D. Glinki. \\
Tel.: +7 (473) 253-73-71; e-mail: moskalefff@gmail.com

\item[Grebennikov Konstantin Vladimirovich:] Postgraduate Student, Department of Industrial Power Engineering, Voronezh State Technological Academy. \\
Tel.: +7 (473) 255-44-66; e-mail: greb86@mail.ru

\item[Shitov Viktor Vasiljevich:] Doctor of Engineering Science, Full Professor, Head of Department of Industrial Power Engineering, Voronezh State Technological Academy. \\
Tel.: +7 (473) 255-44-66; e-mail: svw@list.ru
\end{description}

\Russian
\section*{Сведения об авторах}

\begin{description}
\item[Москалев Павел Валентинович:] кандидат технических наук, доцент кафедры высшей математики и теоретической механики Воронежского государственного аграрного университета имени К.Д. Глинки. \\
Тел.: +7 (473) 253-73-71; e-mail: moskalefff@gmail.com

\item[Гребенников Константин Владимирович:] аспирант кафедры промышленной энергетики Воронежской государственной технологической академии. \\
Тел.: +7 (473) 255-44-66; e-mail: greb86@mail.ru

\item[Шитов Виктор Васильевич:] доктор технических наук, профессор, заведующий кафедрой промышленной энергетики Воронежской государственной технологической академии. \\
Тел.: +7 (473) 255-44-66; e-mail: svw@list.ru
\end{description}


\begin{biblist}[0]{9}\normalsize

\English
\bibitem{moskaleff.2011.01}
\textit{Moskalev P.V., Grebennikov K.V., Shitov V.V.} 
Statistical estimation of percolation cluster parameters 
\Russian [Электронный ресурс] \English // 
arXiv:1105.2334v1 [cond-mat.stat-mech]. 
URL: \url{http://arxiv.org/abs/1105.2334v1} 
\Russian (дата обращения 11.05.2011)

\bibitem{moskaleff.2011.02}
\textit{Москалев П.В., Шитов В.В., Гребенников К.В.} 
О распределении выборочных частот узлов перколяционного кластера // 
Информатика: проблемы, методология, технологии: материалы XI Международной научно-методической конференции. Т.2.
Воронеж: ИПЦ ВГУ, 2011. С.54\--58.

\English
\bibitem{sahimi.1994.01}
\textit{Sahimi M.} 
Application of Percolation Theory. 
Taylor \& Francis: London, 1994. 258 p.

\end{biblist}
\end{document}